\documentclass[vecphys]{article}
\usepackage{lmodern}
\usepackage{amssymb,amsmath}
\usepackage{ifxetex,ifluatex}
\usepackage{fixltx2e} % provides \textsubscript
\ifnum 0\ifxetex 1\fi\ifluatex 1\fi=0 % if pdftex
  \usepackage[T1]{fontenc}
  \usepackage[utf8]{inputenc}
\else % if luatex or xelatex
  \ifxetex
    \usepackage{mathspec}
  \else
    \usepackage{fontspec}
  \fi
  \defaultfontfeatures{Ligatures=TeX,Scale=MatchLowercase}
\fi
\IfFileExists{upquote.sty}{\usepackage{upquote}}{}
\IfFileExists{microtype.sty}{%
\usepackage{microtype}
\UseMicrotypeSet[protrusion]{basicmath} % disable protrusion for tt fonts
}{}
\usepackage{hyperref}
\hypersetup{unicode=true,
            pdftitle={17 The Diffusion of Humans and Cultures in the Course of the Spread of Farming},
            pdfkeywords={Cultural diffusion; demic diffusion; farming; Neolithic},
            pdfborder={0 0 0},
            breaklinks=true}
\usepackage{makeidx}
\makeindex
\usepackage{graphicx,grffile}
\makeatletter
\def\maxwidth{\ifdim\Gin@nat@width>\linewidth\linewidth\else\Gin@nat@width\fi}
\def\maxheight{\ifdim\Gin@nat@height>\textheight\textheight\else\Gin@nat@height\fi}
\makeatother
\setkeys{Gin}{width=\maxwidth,height=\maxheight,keepaspectratio}
\usepackage{amsmath}

\ifx\paragraph\undefined\else
\let\oldparagraph\paragraph
\renewcommand{\paragraph}[1]{\oldparagraph{#1}\mbox{}}
\fi
\ifx\subparagraph\undefined\else
\let\oldsubparagraph\subparagraph
\renewcommand{\subparagraph}[1]{\oldsubparagraph{#1}\mbox{}}
\fi

\title{17 The Diffusion of Humans and Cultures in the Course of the Spread of
Farming}
\providecommand{\subtitle}[1]{}
\subtitle{In: Diffusive Spreading in Nature, Technology and Society, edited by
Armin Bunde, Jürgen Caro, Jörg Kärger, Gero Vogl}
\author{Carsten Lemmen \and Detlef Gronenborn}
\date{\today}

\begin{document}
\maketitle

\newcommand{\refeq}[1]{equation~\ref{#1}}
\newcommand{\refeqs}[2]{equations~\ref{#1}--\ref{#2}}
\newcommand{\reftab}[1]{Table~\ref{#1}}
\newcommand{\refsec}[1]{Section~\ref{#1}}
\newcommand{\reffig}[1]{Figure~\ref{#1}}
\newcommand{\isotope}{\ensuremath{\delta^{18}\mathrm{O}}}
\newcommand{\degree}{\ensuremath{{}^\circ}}
\newcommand{\comment}[1]{}
\newcommand{\yrbp}{cal\,BP}

\newcommand{\emath}[1]{\ensuremath{#1}}
\newcommand{\mtext}[1]{\emath{\mathrm{#1}}}
\newcommand{\T}{\emath{T}}
\newcommand{\N}{\emath{N}}
\newcommand{\density}{\emath{P}}
\newcommand{\npp}{\emath{E_\mtext{NPP}}}
\newcommand{\lae}{\emath{E_\mtext{LAE}}}
\newcommand{\pae}{\emath{E_\mtext{PAE}}}
\newcommand{\gdd}{\emath{E_\mtext{gdd}}}
\newcommand{\cae}{\emath{E_\mtext{CAE}}}
\newcommand{\si}{\emath{s}}
\newcommand{\fep}{\emath{E_\mtext{FEP}}}
\newcommand{\tli}{\emath{E_\mtext{TLI}}}
\newcommand{\rgr}{\emath{r}}

\newcommand{\dd}[2]{\ensuremath{\frac{\mathrm{d}#1}{\mathrm{d}#2}}}
\newcommand{\del}[2]{\ensuremath{\frac{\partial{}#1}{\partial{}#2}}}
\newcommand{\ddt}[1]{\ensuremath{\frac{\mathrm{d}#1}{\mathrm{d}t}}}
\newcommand{\kma}{\ensuremath{\mathrm{km\,a}^{-1}}}
\newcommand{\pskm}{\ensuremath{\mathrm{km}^{-2}}}
\newcommand{\sqkm}{\ensuremath{\mathrm{km}^{2}}}
\newcommand{\science}[2]{\ensuremath{#1\cdot 10^{#2}}}
\newcommand{\m}[1]{\ensuremath{\left\langle #1\right\rangle}}
\newcommand{\s}[1]{\ensuremath{\left\langle {#1}^2\right\rangle}}

Keywords: Cultural diffusion; demic diffusion; farming; Neolithic

\section{Introduction}\label{introduction}

The most profound change in the relationship between humans and their
environment was the introduction of agriculture and pastoralism. With
this millennia lasting economic shift from simple food acquisition to
complex food production humankind paved the way for its grand
transitional process from mobile groups to sedentary villages, towns and
ultimately cities, and from egalitarian bands to chiefdoms and lastly
states. Given this enormous historic impetus, Gordon Childe many years
ago coined the term ``Neolithic Revolution''~{[}1{]}.

The first experiments began with the end of the Glacial period about
10000 years ago in the so called Fertile Crescent~{[}2{]}. They were
followed by other endeavors in various locations both in the Americas
and in Afroeurasia. Today farming has spread to all but the most
secluded or marginal environments of the planet~{[}3{]}. Cultivation of
plants and animals on the global scale appears to have changed energy
and material flows---like greenhouse gas emissions---so fundamentally,
that the term ``early \index{anthropocene}anthropocene'' is considered
for the era following the Mid-Holocene~{[}4{]}.

\hyperdef{}{fig:map}{}
\begin{figure}[htbp]
\centering
\includegraphics{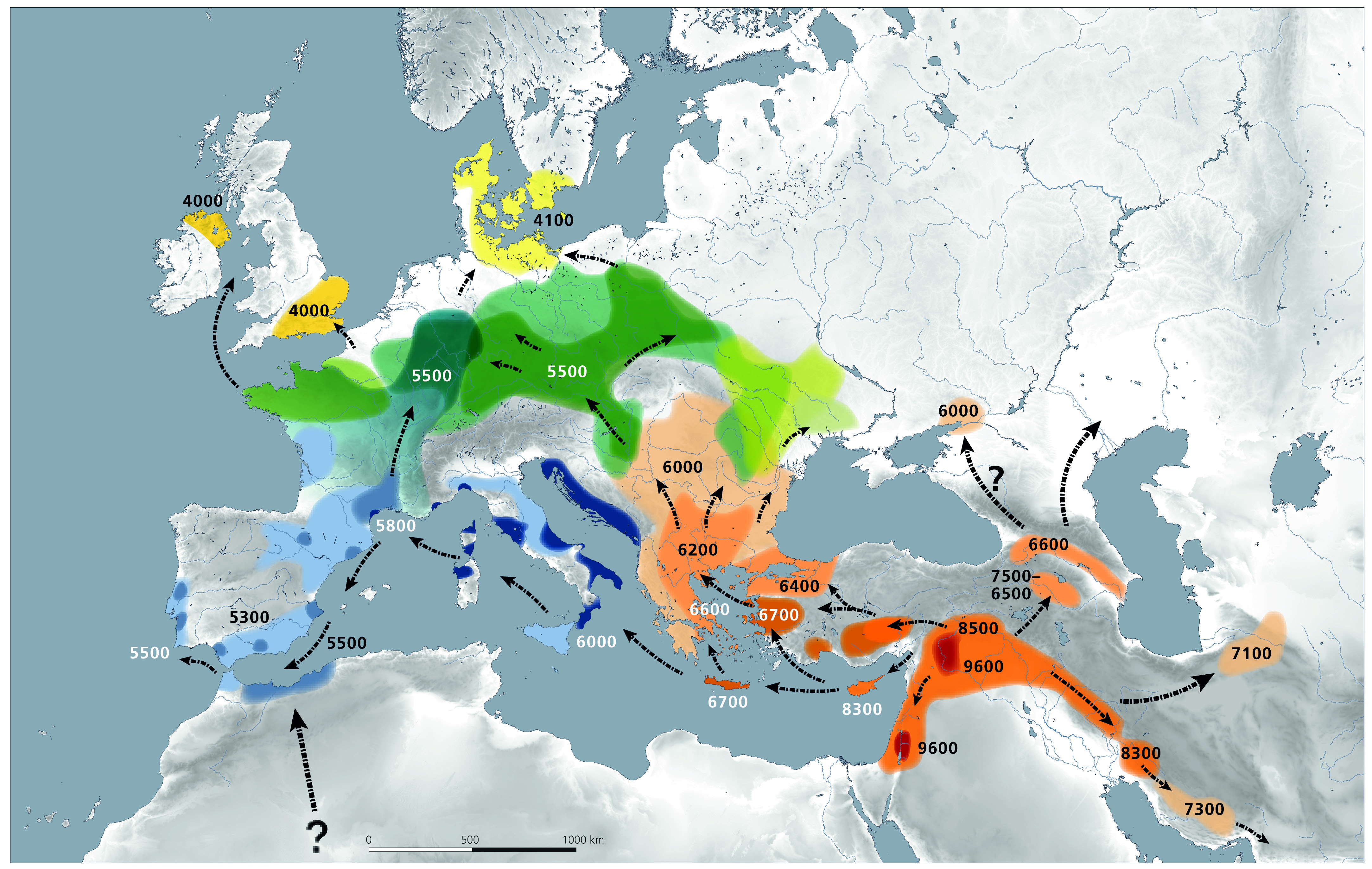}
\caption{Overview of the study area and the archaeologically visible
expansion of farming. \label{fig:map}}
\end{figure}

Possible reasons for the emergence of farming during the relatively
confined period between the Early and Mid-Holocene in locations
independent of each other are continuously being
debated~{[}2{]}~{[}5{]}~{[}6{]}. Once these inventions were in place,
they immediately become visible in the archeological and
paleoenvironmental records. From then on we can trace the spatial
expansion of the newly domesticated plants and animals, the spatial
expansion of a life style based on these domesticates, and the induced
changes in land cover~{[}7{]}~{[}8{]}. From this empirically derived
data the characteristic condensed map of the spread of farming into
western Eurasia is produced (\hyperref[fig:map]{Figure 17.1}).

The local changes introduced spatial differences in knowledge, labor,
technology, materials, population density, and---more
indirectly---social structure and political organization, amongst
others~{[}9{]}~{[}10{]}. Consequently, the dynamics occurring along such
\index{spatiotemporal gradient}spatial gradients may be modeled as a
diffusive process. In Chapter 2, Fick's first law was introduced, which
describes that the average flux across a spatial boundary is
proportional to the difference of concentration across this boundary
(Chapter 2, Eq. 2.6). Each of the local inventions would then spread
outward from its respective point of origin. Indeed, these
\index{spatiotemporal gradient}spatiotemporal gradients have been
observed in ceramics~{[}1{]}, radiocarbon dates~{[}7{]}~{[}11{]},
domesticates~{[}12{]}~{[}13{]}, land use change~{[}14{]}~{[}15{]}, and
the genetic composition of
paleopopulations~{[}16{]}~{[}17{]}~{[}18{]}~{[}19{]}.

For an understanding of the expansion process, it appears appropriate to
apply a diffusive model. Broadly, these numerical modeling approaches
can be categorized in correlative, continuous and discrete. Common to
all approaches is the comparison to collections of radiocarbon data that
show the apparent \index{wave of advance}wave of advance of the
transition to farming. However, these data sets differ in entry density
and data quality. Often they disregard local and regional specifics and
research gaps, or dating uncertainties. Thus, most of these data bases
may only be used on a very general, broad scale. One of the pitfalls of
using irregularly spaced or irregularly documented radiocarbon data
becomes evident from the map generated by Fort (this volume, Chapter
16): while the general east-west and south-north trends become evident,
some areas appear as having undergone anomalously early transitions to
farming. This may be due to faulty entries into the data base or
regional problems with radiocarbon dating, if not unnoticed or
undocumented laboratory mistakes.

\textbf{Correlative models} compare the timing of the transition (or
other archeologically visible frontier) with the distance from one or
more points of origin. These are among the earliest models proposed,
such as those by Clark~{[}7{]} or by Ammerman~{[}20{]}. These models
have been used to roughly estimate the front propagation speed of the
introduction of agriculture into Europe, and the original speed of
around 1 km per year has not been substantially refined until today.

\textbf{Continuous models} predict at each location within the specified
domain the transition time as the solution of a differential equation,
mostly of a Fisher--Skellam type, in relation to the distance from one
or more points of origin. Often this distance is not only the geometric
distance but also factors in geography, topography and even ease of
migration. The prediction from the continuous model is compared to the
archeologically visible frontier~{[}21{]}. This is the approach taken by
Fort (this volume, Chapter 16) who compares the wave-front propagation
of different models for the transition from a hunting and gathering
economy to a farming economy in Europe with the
\index{spatiotemporal gradient}spatiotemporal pattern of the earliest
radiocarbon dates locally associated with farming.

\textbf{Discrete models} Discrete models are realized as agent-based
models (see also Chapter 2.5), with geographic areas (or their
populations) representing agents, and rules that describe the
interaction, especially the diffusion properties, between them. They
also predict for each geographic area the transition time, but not as an
analytic, but rather as an emergent property of the system. We here
introduce as an example the discrete agent-based and gradient-adaptive
model the Global Land Use and technological Evolution Simulator (GLUES).

\section{Agent-based gradient-adaptive model
GLUES}\label{agent-based-gradient-adaptive-model-glues}

\hyperdef{}{fig:cells}{}
\begin{figure}[htbp]
\centering
\includegraphics{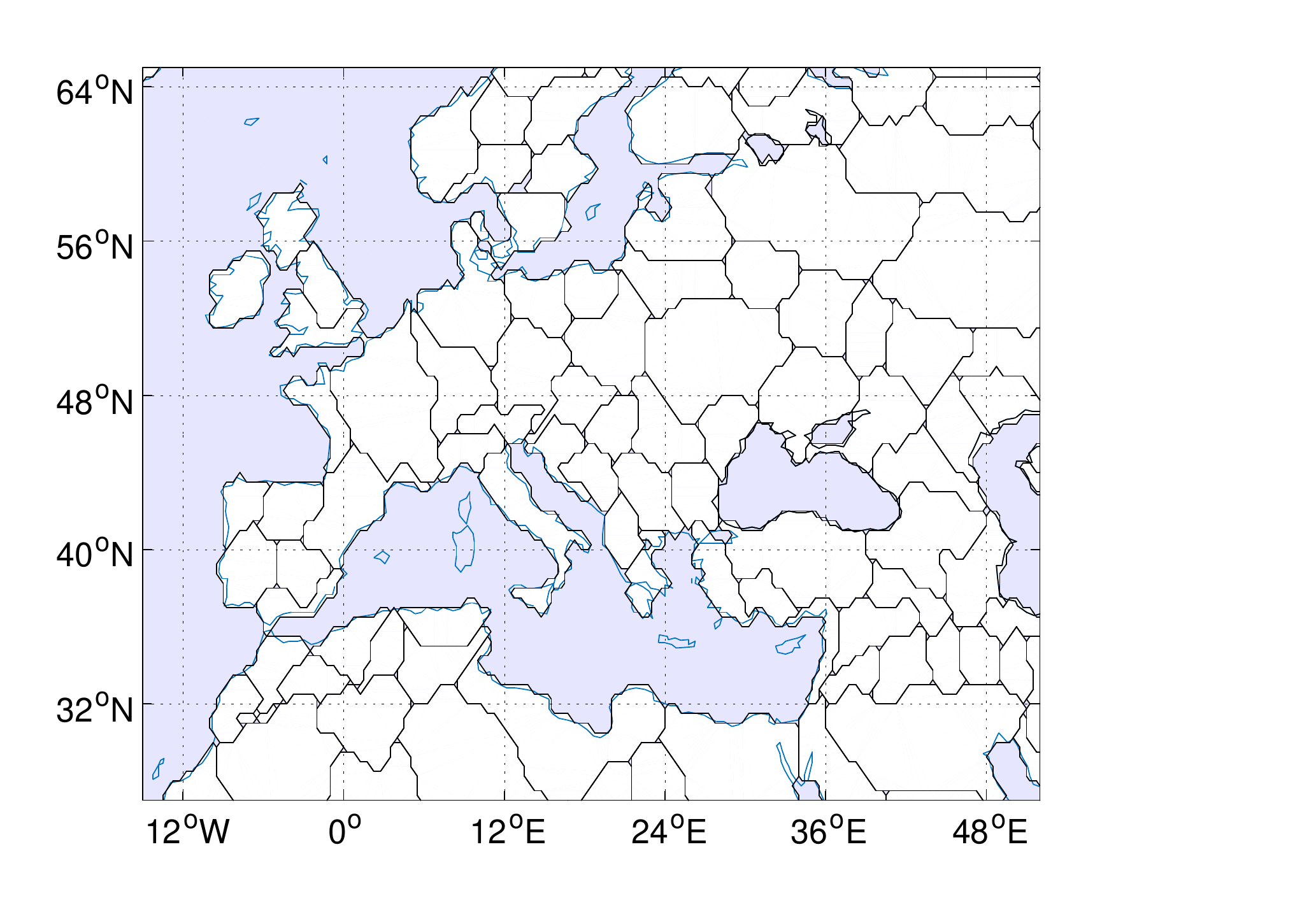}
\caption{Regions constituting the set of agents in the simulation (shown
for Western Eurasia and North Africa) in 685 globally distributed
regions) \label{fig:cells}}
\end{figure}

We employ the Global Land Use and technological Evolution Simulator
(GLUES~{[}22{]})--a numerical model of prehistoric innovation,
demography, and subsistence economy based on interacting geographic
populations as agents and gradient adaptive trait dynamics to describe
local evolution. There are currently 685 regions representing the
``cells'' of agent-based models (\hyperref[fig:cells]{Figure 17.2}),
together with interaction rules that describe diffusion of people,
material and information between these regions. The agent is the
population living within a region, and its state is described by its
density and three characteristic population-average trait
\index{characteristic traits} that evolve towards optimal local growth
rate.

This numerical model is able to hindcast the regional transitions to
agropastoralism and the diffusion of people and innovations across the
world for the time span between approximately 8000~BCE (before the
common era) and 1500~CE. It has been successfully compared to
radiocarbon data for Europe~{[}22{]}, Eastern North America~{[}23{]},
and South Asia~{[}24{]}.

Regions are generated from ecozone clusters that have been derived to
represent homogeneous net primary productivity (\npp) based on a 3000
BCE 1\degree{} \(\times\) 1\degree{} paleoproductivity estimate; this
estimate was derived from a climatologically downscaled dynamic
paleovegetation simulation~{[}25{]}. By using \npp, many of the
environmental factors taken into account by other expansion or
predictive models, such as altitude, latitude, rainfall, or
temperature~{[}12{]}~{[}26{]} are implicitly considered.

\subsection{Local characteristic
variables}\label{local-characteristic-variables}

In each agent population, the mathematical model resolves the change
over time of population density and three characteristic---meaning
important---population-averaged sociocultural
traits\index{characteristic traits} \(X \in \left\{\T,\N,C\right\}\):
technology \(\T\), share of agropastoral activities \(C\), and economic
diversity \(\N\). They are interpreted as follows:

\begin{enumerate}
\def\labelenumi{\arabic{enumi}.}
\item
  Technology \(\T\) is a trait which describes the efficiency for
  enhancing biological growth rates, or diminishing mortality. It is
  represented by the efficiency of food procurement---related to both
  foraging and farming---and improvements in health care. In particular,
  technology as a model describes the availability of tools, weapons,
  and transport or storage facilities, and includes institutional
  aspects related to work organization and knowledge management. These
  are often synergistic: the technical and societal skill of writing as
  a means for cultural storage and administration, with the latter
  acting as an organizational lubricant for food procurement and its
  optimal allocation in space and among social groups;
\item
  Economic diversity \(\N\) resolves the number of different
  agropastoral economies available to a regional population. This trait
  is closely tied to regional vegetation resources and climate
  constraints. A larger economic diversity offering different niches for
  agricultural or pastoral practices enhances the reliability of
  subsistence and the efficacy in exploiting heterogeneous landscapes.
\item
  A third model variable \(C\) represents the share of farming and
  herding activities, encompassing both animal husbandry and plant
  cultivation. It describes the allocation of energy, time, or manpower
  to agropastoralism with respect to the total food sector; this is the
  only variable that is directly comparable to data from the
  archeological record.
\end{enumerate}

\begin{table*}
\centering
\small
\begin{tabular}{|l l|l r|}\hline
Characteristic trait\index{characteristic traits} & Symbol & Quantification & Typical range\\ \hline
Technology efficiency & $\T$ & Factor of efficiency gain over Mesolithic & $0.9$--$15$\\
Economic diversity & $\N$ & Richness of economic agropastoral strategies & $0.1$--$8$\\
Agropastoral share & $C$ & Fraction of activities in agropastoralism & $0$--$1$\\
\hline
\end{tabular}
\caption{Characteristic traits\index{characteristic traits} used in the Gradient Adaptive Dynamics\index{adaptive dynamics} formulation of GLUES; a full
table of symbols used is available as \reftab{tab:symbols}.}
\label{tab:traits}
\end{table*}

\subsection{Adaptive dynamics}\label{adaptive-dynamics}

The adaptive coevolution of the food production system \(\{\T,\N,C\}\)
and population density \(P\) follow the conceptual model that was, e.g,
proposed by Boserup~{[}27{]}: ``The close relationship which exists
today between population density and food production system is the
result of two long-existing processes of adaptation. On the one hand,
population density has adapted to the natural conditions for food
production {[}{]}; on the other hand, food supply systems have adapted
to changes in population density.''

Mathematically, this conceptual model is implemented in the Gradient
Adaptive Dynamics\index{adaptive dynamics} (GAD) approach: Whenever
traits can be related to growth rate, then an approach known as adaptive
dynamics can be applied to generate the equations for the temporal
change of traits, the so-called evolution equations. This adaptive
dynamics\index{adaptive dynamics} was introduced by Metz~{[}28{]} and
goes back to earlier work by Fisher in the 1930s~{[}29{]} and the field
of genetics. When genetically encoded traits influence the
fitness\index{fitness gradient} of individuals, that gene's prevalence
in a population changes. Adaptive dynamics\index{adaptive dynamics}
describes the change of the probability of the trait in the population
by considering its mutation rate and its fitness
gradient\index{fitness gradient}, i.e., the marginal
benefit\index{marginal benefit} of changes in the
trait\index{characteristic traits} for the (reproductive)
\index{fitness gradient}fitness of the individual.

To ecological systems, this metaphor was first applied in 1996~{[}30{]},
and to cultural traits in 2003~{[}25{]}; in this translation, the
genetically motivated term mutation rate was replaced by the
ecologically observable variability of a trait. Because many traits are
usually involved in (socio)ecological applications (here \(\T, \N, C\)),
the term Gradient Adaptive Dynamics\index{adaptive dynamics} was
introduced to emphasize the usage of the growth-rate gradient of the
vector of traits.

\subsubsection{Aggregation}\label{aggregation}

Gradient Adaptive Dynamics\index{adaptive dynamics} describes the
evolution of one or more aggregated (population-average) characteristic
traits\index{characteristic traits}. By definition, it requires
variability within a population, and is thus suitable for the
description of medium-size to large populations.

In a local population \(B\) composed of \(n\) sub-population members
\(\iota \in \{1\ldots n\}\), each member with relative contribution
\(B_\iota/B\), characteristic traits\index{characteristic traits}
\(X_\iota\), and time-dependent environmental condition \(E_\iota(t)\),
has a relative growth rate \(r_\iota\)

\begin{equation}\label{eq:growthsize}
r_\iota=\frac{1}{B} \cdot \ddt{B_\iota} = r_\iota \left(X_\iota,E_\iota(t)\right).
\end{equation}

This equation is often formulated in terms of the population density
\(P = B / A\), where \(A\) is the area populated by \(B\):

\begin{eqnarray}\label{eq:growthdensity}
r_\iota = \frac{1}{P} \cdot \ddt{P_\iota} & = & r_\iota \left(X_\iota,E_\iota(t)\right)\\
\textrm{and}\qquad\sum_\iota^n (P_\iota/P) & = & 1.\nonumber
\end{eqnarray}

The mean of a quantity \(X\) over all individuals \(\iota\) is
calculated as

\begin{equation}
\m{X}=\sum_{\iota=1}^n \frac{X_\iota P_\iota}{P}.
\label{eq:mX}
\end{equation}

The adaptive dynamics\index{adaptive dynamics} rooted in genetics
assumes that mutation errors are only relevant at cell duplication, and
not during cell growth. Translated to the ecological entity population
this restriction enforces that all traits \(X_\iota\) of a member of
this population are stable during the lifetime of this member:
\(\ddt{}X_\iota=0\) for all \(X_\iota\). Changes in the aggregated
traits \(\m{X}\) are a result of frequency selection (the number of
members carrying a specific characteristic
trait\index{characteristic traits} increases or decreases as a result of
selection) only.

\subsubsection{Aggregated trait
dynamics}\label{aggregated-trait-dynamics}

Aggregation of Eq. \ref{eq:mX}, differentiation with respect to time,
and considering \(\ddt{}X_\iota=0\), gives

\begin{eqnarray}
\ddt{\m{X}} & = & \sum_{\iota=1}^n \del{\frac{X_\iota P_\iota}{P}}{t} \nonumber\\
& = & \m{X\cdot r(X)} - \m{r(X)}\cdot\m{X} \label{eq:diff}
\end{eqnarray}

Using a Taylor expansion of \(r\) about \(X=\m{X}\), Eq. \ref{eq:diff}
can be reformulated in terms of the \((k+1)^\mathrm{st}\) central
moment, of which the first summand is zero. Neglecting moments higher
than \(k=2\), the temporal change of \m{X} is

\begin{equation}
\ddt{\m{X}}
  =\sigma_X^2\cdot \del{}{X}r\left(\m{X}\right)
  + v_X\sigma_X^3 \del{^2}{X^2}r\left(\m{X}\right),
\label{eq:gad}
\end{equation}

where \(\sigma_X^2=\m{(X-\m{X})^2}\) denotes the variance of \(X\), and
\(v_X\) describes the skewness of \(X\). Essentially, the population
averaged value of a trait changes at a rate which is proportional to the
\index{marginal benefit}marginal benefit
(\index{fitness gradient}fitness) induced by trait changes on the growth
rate \(r\) evaluated at the mean
\index{characteristic traits}characteristic trait value.

If the probability distribution of \(X\) is known, the variance and
skewness can be deduced; in other cases, variance and skewness have to
be specified explicitly. The third moment may not be necessary in all
cases and the approximation can be truncated after the second order
term; if not, specific closure terms have to be derived. The use of the
partial derivative on the right hand side of Equation (\ref{eq:gad})
reflects that in all applications of gradient \index{adaptive dynamics}
so far, non-local effects have been disregarded.

\begin{figure}[htbp]
\centering
\includegraphics{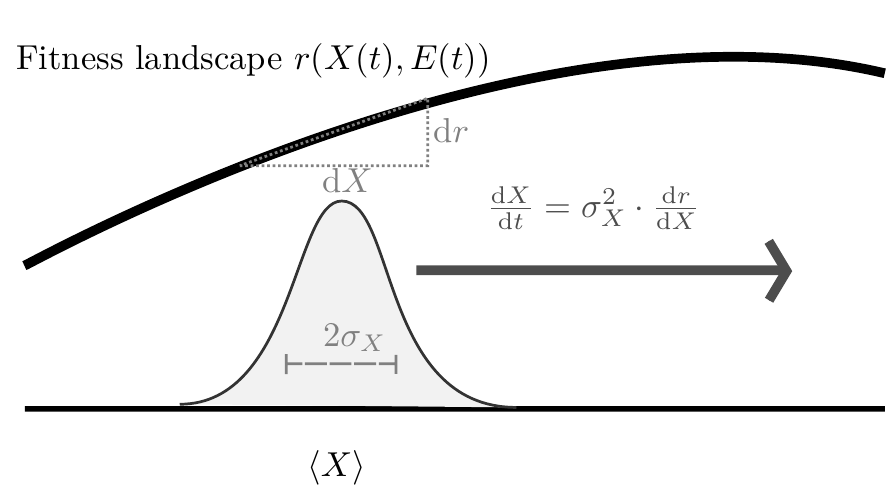}
\caption{The adaptive dynamics\index{adaptive dynamics} of a
characteristic trait\index{characteristic traits} \(X\) in a
\index{fitness landscape}fitness landscape \(r(X,E(t))\) is described by
the width of the trait distribution (\(\sigma_X\)) and the
\index{marginal benefit}marginal benefit that a small change in \(X\)
has on the growth rate \(\rgr\). Modified from~{[}31{]}.\label{fig:gad}}
\end{figure}

Leaving out the angular brackets around \(X\) for better readability,
and truncating at \(k=1\), the gradient
\index{adaptive dynamics}adaptive dynamics for
\index{characteristic traits}trait \(X\) is given by

\begin{equation}
\ddt{X} =\sigma_X^2\cdot \del{r(X)}{X}
\label{eq:gadsimple}
\end{equation}

with \(X \in \{\T,\N,C\}\). Equation (\ref{eq:gadsimple}) is visually
shown in \hyperref[fig:gad]{Figure 17.3}.

\hyperdef{}{fig:gad}{}

\subsection{Local population growth}\label{local-population-growth}

Key to \index{adaptive dynamics}adaptive dynamics is the formulation of
the growth rate as a function of all
\index{characteristic traits}characteristic traits. Once this dependence
is specified, the evolution equations for \(X\) are generated
automatically from Eq. \ref{eq:gadsimple}.

\begin{table*}
\centering
\small
\begin{tabular}{|l l|l l|}\hline
Symbol & description & unit & typical range \\ \hline
$P$ & population density & \pskm & $>0$\\
$X$ & growth-influencing trait &  & $>0$\\
$t$ & time & a & 9500--1000~BCE\\
$r$  & specific growth rate & a$^{-1}$ & \\
$E$ & environmental constraints && \\
$\m{\cdot}$ & mean / first moment of $\cdot$ & & \\
$\sigma^2$   & variance & & $>0$ \\
$\varsigma$ & diffusion parameter & & $>0$\\
$v$     & skewness && \\ \hline
$\T$ & technology trait & & $>0$\\
$\N$ & economic trait & & $>0$\\
$C$ & labor allocation trait & & $]0;1[$\\
\si & subsistence intensity &&\\
\tli & temperature limitation && $[0;1]$\\
\fep & food extraction potential&& $[0;1]$\\
$\omega$ & administration parameter && \\
$\gamma$ & exploitation parameter && \\
$\mu$ & fertility rate & a$^{-1}$ &\\
$\rho$ & mortality rate & a$^{-1}$&\\ \hline
\end{tabular}
\caption{Symbols and variables used in the text and equations (in order of appearance in text).
A useful parameter set is
$\mu=\rho=0.004$,
$\omega=0.04$,
$\gamma=0.12$,
$\delta_{\T}=0.025$,
$\varsigma_{\mathrm{demic}}=0.002$,
$\varsigma_\mathrm{info}=0.2$,
$\delta_{\N}=0.9$;  a variable $\delta_{C}=C\cdot(1-C)$; and initial values for $P_{0}=0.01$, $\T_{0}=1.0$, $\N_{0}=0.8$, and $C_{0}=0.04$.
}
\label{tab:symbols}
\end{table*}

The relative growth rate \(\rgr\) of an agent population with density
\(\density\) is the sum of gain and loss (often termed birth and death
rates, respectively; see \hyperref[tab:symbols]{Table 1} for an overview
of symbols used)

\begin{equation}
\rgr = \mu \cdot \textrm{gain} - \rho \cdot \textrm{loss},
\end{equation}

with loss and gain coefficients \(\mu\) and \(\rho\).

Out of several factors contributing to population gain, the Neolithic
transition is foremost characterized by changes in subsistence intensity
(\si). Subsistence intensity describes a community's effectiveness in
generating consumable food and secondary products; this can be achieved
based on an agricultural (with fractional activity \(C\)) and a
hunting-gathering life style (with fractional activity \(1-C\)). The
quantity \({\si}\) is dimensionless and scaled such that a value of
unity expresses the mean subsistence intensity of a hunter-gatherer
society equipped with tools typical for the mature Mesolithic
(\(\T = 1, C \approx 0\)).

\begin{equation}
\label{eq:food}
{\si}= (1-C)\cdot\sqrt{\T} +  C\cdot\N\cdot\T\cdot\tli
\end{equation}

The agropastoral part of \si{} increases linearly with \(\N\) and with
\(\T\): The more economies (\N) there are, the better are sub-regional
scaled niches utilized and the more reliable returns are generated when
annual weather conditions are variable; the higher the technology level
(\T), the better the efficiency of using natural resources (by
definition of \T). While a variety of techniques can steeply increase
harvests of domesticated species, analogous benefits for foraging
productivity are less pronounced and justify a less than linear
dependence of the hunting-gathering calorie procurement on \(\T\), such
as a square root formulation.

We introduce an additional temperature constraint (\(\tli\)) on
agricultural productivity which considers that cold temperature could
only moderately be overcome by Neolithic technologies. This limitation
is unity at low latitudes and approaches zero at permafrost conditions.

The domestication process is represented by \(\N\), which is the number
of realized agropastoral economies. We link \(\N\) to natural resources
by expressing it as the fraction \(f\) of potentially available
economies (\(\pae\)) by specifying \(\N=f\cdot\pae\), where the latter
corresponds to the richness in domesticable animal or plant species
within a specific region.

To account for overexploitation of natural resources and productivity
losses by societal organization, \({\si}\) is multiplied by two terms
representing those processes. \((\fep-\gamma\sqrt{\T}\density)\)
expresses the effects of overexploitation of natural resources \(\fep\)
by increasing impact, calculated by the (scaled) product of technology
and density ; organizational losses within a society are expressed by
the term \((1-\omega\,\T)\): as technology advances, more and more
people neither farm nor hunt: Construction, maintenance, administration
draw a small fraction \(\omega\) of the workforce away from
food-production. Gathering those gain terms, the growth rate equation is

\begin{equation}
r=\mu \cdot (\fep-\gamma\sqrt{\T}\density) \cdot (1-\omega\,\T)
\cdot {\si}-\textrm{loss}
\end{equation}

The loss term takes a standard ecological form modeled on the crowding
effect (also known as ecological capacity), and is thus proportional to
\(\density\). It is mediated by technologies (\T, with
\(\T_\mathrm{lit}=12\)), which mitigate, for example, losses due to
disease. The full growth rate equation is then:

\begin{equation}
r=\mu \cdot (\fep-\gamma\sqrt{\T}\density) \cdot (1-\omega\,\T)
\cdot {\si}-\rho\cdot\density\cdot\mathrm{e}^{-\T/\T_{\mathrm{lit}}},
\end{equation}

\subsection{Spatial diffusion model}\label{spatial-diffusion-model}

\hyperdef{}{fig:flux}{}
\begin{figure}[htbp]
\centering
\includegraphics{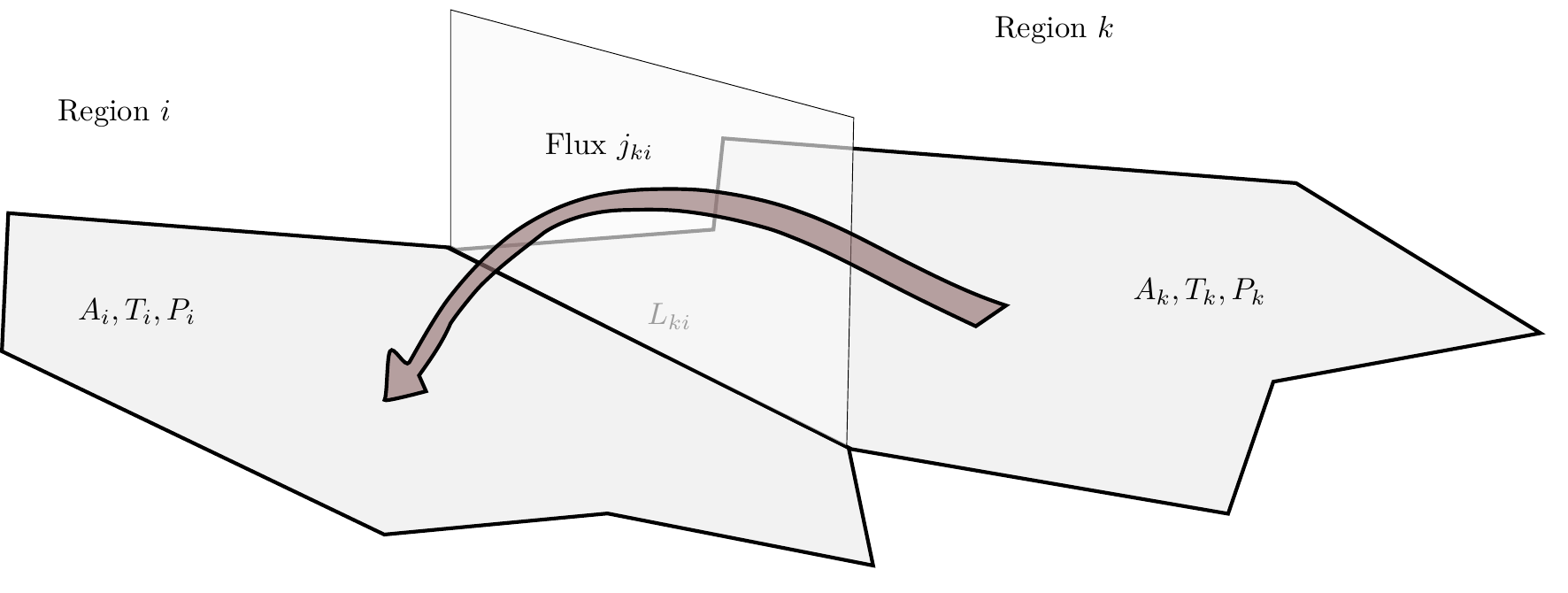}
\caption{Schematic representation of interregional exchange in
GLUES\label{fig:flux}.}
\end{figure}

Information, material and people are diffused between regions with a
flux modeled on Fickian diffusion (Chapter 2, eqs 2.6--2.9), modified
for the discrete two-dimensional region arrangement and with a locally
heterogeneous diffusion coefficient \(D_{ik}\). The change of any
characteristic trait\index{characteristic traits} \(X_i\) in a region
\(i\) due to diffusion from/to all regions \(k \in \mathcal{N}_k\) in
its neighborhood \(\mathcal{N}_i\) with neighbor distance
\(\Delta x_{ik}\) is

\begin{equation}
\displaystyle{\frac{\Delta X_{i}}{\Delta t}}
=  \sum_{k\in\mathcal{N}_{i}} - j_{ik} / \Delta x_{ik},
\label{eq:fickdiscrete}
\end{equation}

with \(j_{ik} = -D_{ik} \Delta X_{ik} / \Delta x_{ik}\) constituting the
diffusive flux between \(i\) and \(k\) (\hyperref[fig:flux]{Figure
17.4}); This gives

\begin{equation}
\displaystyle{\frac{\Delta X_{i}}{\Delta t}}
=  \sum_{k\in\mathcal{N}_{i}}
D_{ik}  \Delta X_{ik}  \Delta x_{ik}^{-2}.
\label{eq:fickdiscretefull}
\end{equation}

This equation can be reformulated~{[}32{]} as

\begin{equation}
\displaystyle{\frac{\Delta X_{i}}{\Delta t}} =
\varsigma \sum_{k\in\mathcal{N}_{i}} f_{ik}  \Delta X_{ik}
\label{eq:fickold}
\end{equation}

with \(f_{ik} = j_{ik} \Delta x_{ik}^{-2} \varsigma^{-1}\), where
\(\varsigma\) is a global diffusion property characterizing the
underlying process (see below) and \(f_{ik}\) collects all regionally
varying spatial and social diffusive aspects.

The social factor in the formulation of \(f_{ik}\) is the difference
between two regions' influences , where influence is defined as the
product of population density \(P\) and technology \(T\), scaled by the
average influence of regions \(i, k\). The geographic factor is the
conductance between the two regions, which is constructed from the
common boundary length \(L_{ik}\) divided by the mean area of the
regions \(\sqrt{A_i A_k}\). Non-neighbour regions by definition have no
common boundary, and hence have zero conductance; to connect across the
Strait of Gibraltar, the English Channel, and the Bosporus, the
respective conductances were calculated as if a narrow land bridge
connected them. No additional account is made for increased conductivity
along rivers~{[}33{]}, as the regional setup of the model is biased
(through the use of \npp{} similarity clusters) toward elongating
regions in the direction of rivers. Altitude and latitude effects are
likewise implicitly accounted for by the \npp{} clustering in the region
generation.

\subsection{Three types of diffusion}\label{three-types-of-diffusion}

Three types of diffusion are distinguished: (1) demic
diffusion\index{demic diffusion}, i.e.~the migration of people, (2) the
hitchhiking of traits with migrants, and (3) cultural
diffusion\index{cultural diffusion}, i.e.~the information exchange of
characteristic traits\index{characteristic traits}.

\index{demic diffusion}\textbf{Demic diffusion} is the mass-balanced
migration of people between different regions. The diffusion equation
\ref{eq:fickold} is applied to the number of inhabitants \(B_i=P_i A_i\)
in each region \(i\).

\begin{equation}
\left.\displaystyle{\frac{\mathrm{d}B_{i}}{\mathrm{d}t}}\right\vert_{\mathrm{demic}}
= \begin{cases}
\varsigma_\mathrm{demic} \sum_{j\in\mathcal{N}_i} f_{ij} (B_j - B_i), & \rgr \ge 0\\
0 & \text{otherwise}
\end{cases}\label{eq:dXdtdemic}
\end{equation}

\noindent The free parameter \(\varsigma_{\mathrm{demic}}\) has to be
determined from comparison to data. The parameter estimation based on
the European dataset by Pinhasi~{[}34{]} and the typical front speed
extracted from this dataset yields \(\varsigma_{\mathrm{demic}}=0.002\)
(see~{[}32{]} for parameter estimation). We impose an additional
restriction to migration by requiring positive growth rate
\(r_i \ge 0\), i.e.~favorable living conditions, in the receiving region
\(i\).

\textbf{Hitchhiking traits}: Whenever people move in a demic process,
they carry along their traits to the receiving region:

\begin{equation}
\left.\displaystyle{\frac{\mathrm{d}X_{i}}{\mathrm{d}t}}\right\vert_{\mathrm{demic}}
= \begin{cases}
\varsigma_\mathrm{demic} \sum_{j\in\mathcal{N}_{i},f_{ij}>0} f_{ij}X_{j}\frac{B_j}{B_i}, & \rgr \ge 0\\
0 & \text{otherwise}
\end{cases}\label{eq:dXdthitch}
\end{equation}

\textbf{Information exchange}: Traits do not decrease when they are
exported. Thus, only the positive contribution from the diffusion
equation \ref{eq:fickold} is considered, information exchange is not
mass-conserving.

\begin{equation}
\left.\displaystyle{\frac{\mathrm{d}X_{i}}{\mathrm{d}t}}\right\vert_{\mathrm{info}} =
\varsigma_\mathrm{info} \sum_{j\in\mathcal{N}_{i},f_{ij}>0} f_{ij}\cdot(X_{j}-X_{i}) \label{eq:dXdtinfo}
\end{equation}

\noindent The diffusion parameter was estimated to be
\(\varsigma_{\mathrm{info}}=0.2\) in a reference scenario. Despite the
formal similarity of Eqs. (\ref{eq:dXdtinfo}, \ref{eq:dXdtdemic}),
suggesting a mere factor
\(\varsigma_\text{demic} / \varsigma_\text{info}\) as the difference,
the processes are rather different: migration is mass-conserving,
information exchange is not (note the summation of only positive
\(f_{ij}\) for information exchange); migration is hindered by bad
living conditions, information exchange is not.

\subsection{Availability and reproduction of
results}\label{availability-and-reproduction-of-results}

The numerical model and necessary datasets have been publicly released
under an open source license. The code is available from SourceForge
are given in \hyperref[tab:symbols]{Table 1}

\section{Model applications to diffusion
questions}\label{model-applications-to-diffusion-questions}

Two questions have been addressed with GLUES that are specific to
diffusion. First and foremost, the wave front propagation speed was
diagnosed from the model with respect to both demic and cultural
diffusion~{[}22{]}. For a mixed demic\index{demic diffusion} and
cultural diffusion\index{cultural diffusion} scenario, the authors found
a wave front propagation speed of \(0.81\) km a\textsuperscript{-1}
radiating outward of an assumed center near Beirut (Lebanon) in the
European dataset, somewhat faster than the speed diagnosed from
radiocarbon data (\(0.72\) km a\textsuperscript{-1}~{[}34{]}). Both in
the radiocarbon data and the model simulation, however, there is large
scatter from the linear time-distance relationship, with a lower than
average propagation speed in the Levante before 7000~BCE, and with
higher than average propagation speed with the expansion of the
Linearbandkeramik in the 6th millennium~BCE.

It was also found, that there is a regionally heterogeneous contribution
of demic and cultural diffusion, and of local innovation in the
simulated transition to agropastoralism. While either diffusion
mechanism is necessary for a good reconstruction of the emergence of
farming, the major contribution to local increases in \(\T\) or \(C\) is
local innovation. Diffusion (its contribution is in many regions around
20\% to the change in an effective variable) seems to have been a
necessary trigger to local invention.

Not only is the contribution of diffusive processes heterogeneous in
space, but it also varies in time. This was shown by studying the
interregional exchange fluxes in the transition to farming for Eurasia
with GLUES~{[}32{]}. Most Eurasian regions exhibited an equal proportion
of demic and cultural diffusion events when integrated over time, with
the exception of some mountainous regions (Alps, Himalayas), where demic
diffusion\index{demic diffusion} is probably overestimated by the model:
the higher populations in the surrounding regions may lead to a constant
influx of people into the enclosed and sparsely inhabited mountain
region.

When time is considered, however, it appears that diffusion from the
Fertile Crescent is predominantly demic before 4900~BCE, and cultural
thereafter; that east of the Black Sea, diffusion is demic until 4200
BCE, and cultural from 4000~BCE. The expansion of Southeastern and
Anatolian agropastoralism northward is predominantly cultural at 5500
BCE, and predominantly demic 500 years later. At 5000~BCE, it is demic
west of the Black Sea and cultural east of the Black Sea; at 4500~BCE,
demic processes again take over part of the eastern Black Sea northward
expansion. This underlines that ``Previous attempts to prove either
demic\index{demic diffusion} or cultural
diffusion\index{cultural diffusion} processes as solely responsible
{[}..{]} seem too short-fetched, when the spatial and temporal
interference of cultural and diffusive processes might have left a
complex imprint on the genetic, linguistic and artefactual
record''~{[}32{]}.

\hyperdef{}{fig:network}{}
\begin{figure}[htbp]
\centering
\includegraphics{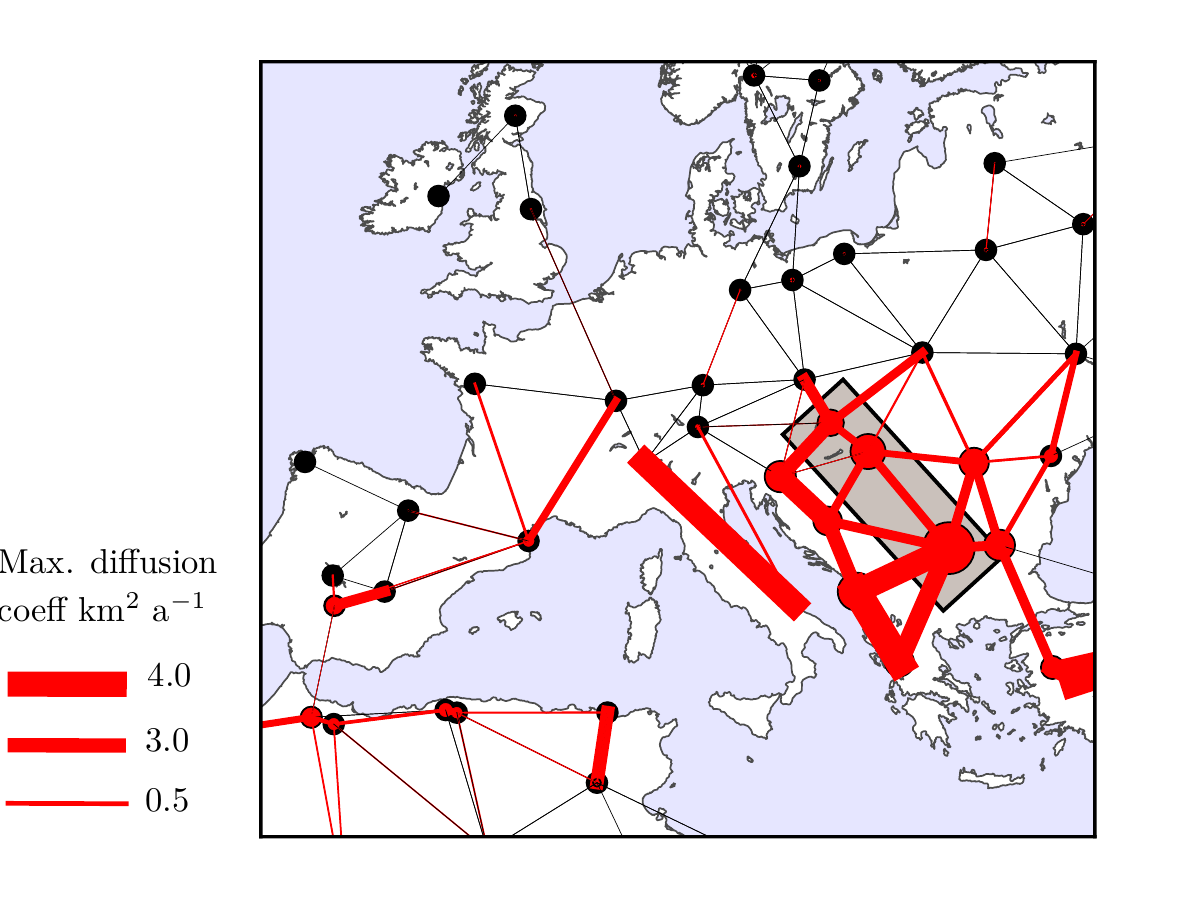}
\caption{Topology of European regional connections and maximum diffusion
coefficient for each region. Circles represent geographic centers of
regions, red circles highlight regions with large maximum influence; the
size of the highlighted connections represents the maximum diffusion
coefficient between two adjacent region. Shading indicates the three
regions analysed in \hyperref[fig:trajectory]{Figure 17.6}, labelled
Bulgaria, Serbia, Hungary (from south to north). \label{fig:network}}
\end{figure}

Unlike in many other models, the diffusion coefficient \(D\) here is an
emergent property, that varies in space and time, and that varies among
all neighbors of each region. The diffusion coefficient varies between
zero and 7 km\textsuperscript{2} a\textsuperscript{-1};
\hyperref[fig:network]{Figure 17.5} shows the topology of the
interregional connections in Europe and their maximum diffusion
coefficients. Maximum diffusion is highest on the Balkan and within
Italy (up to 4 km\textsuperscript{2} a\textsuperscript{-1}), it is one
order of magnitude lower for all of Northern Europe. This shows the
importance of the Balkans as a central hub for the diffusion of
Neolithic technology, people, and ideas; there seem to have been main
routes for Neolithic diffusion across the central Balkan, along Adriatic
coastlines, or, to a lesser extent, up the Rh\^{}one valley.

The diffusion coefficient \(D\) seems first and foremost to match the
migration rate of populations of ultimately Anatolian/Near Eastern
ancestry into and within Europe. On a continental scale this rate should
have been higher in Southeastern Europe and possibly in Italy, equally
along the Rhône. This is supported by recent archaeological and
archaeogenetic data, at least for Southeastern and Central Europe
~{[}18{]}~{[}19{]}. Therefore, it is to be assumed that the proportion
of non-indigenous populations should have been highest in these areas.
Towards the north the spread of these immigrant Neolithic populations
was halted until about 4000~BCE, after which farming spread further into
the European north and northwest as well as to the British Isles. This
stagnation pattern is visible from archaeological
evidence~{[}15{]}~{[}35{]} and represented in model
simulations~{[}36{]}. Towards the continental west the evidence for a
lesser proportion of allochtonous cultural traits in the archeological
record of farming societies has continuously been interpreted as an
increase in indigenous populations within these societies; therefore the
rate of immigrants should have been lower. This has at least been
suggested by archaeology~{[}37{]}; recent genetic studies have shown,
however, that the influx of a population of ultimately
Balkanic/Anatolian origin seems also to have been strong at least in the
Paris Basin and eastern France~{[}38{]}.

While the simulated Neolithic transition is reasonably well reflected on
the continental scale, the model skill in representing the individual
regional spatial expansions varies. For example, the particular
geographic expansion of the LBK in Central Europe occurs too late and is
too small in extent towards the Paris Basin. On the other hand, the
timing of the arrival of the Neolithic in the Balkans, in southern
Spain, or in northern Europe is well represented~{[}36{]}.

\hyperdef{}{trajectory}{}
\begin{figure}[htbp]
\centering
\includegraphics{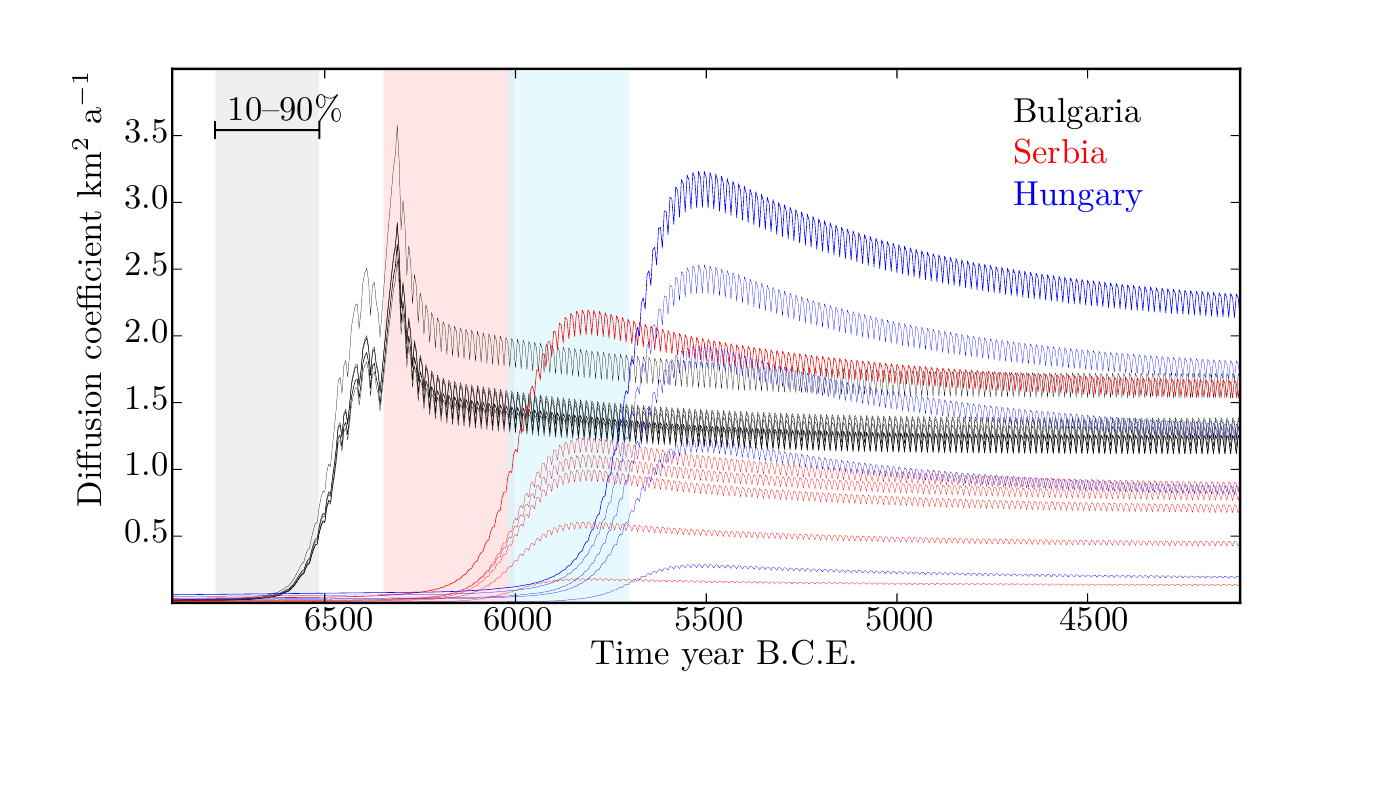}
\caption{Time evolution of the diffusion coefficient for three selected
regions of central Europe (Bulgaria, Serbia, Hungary; shown in black,
red, and blue, respectively). For each of these three regions, the
family of trajectories represents the region's diffusion coefficients
with respect to each of its adjacent regions. The shaded bars indicate
the time interval of a regional transition to agropastoralism in the
simulation (10--90\% of \(C\)).\label{fig:trajectory}}
\end{figure}

For three selected regions along the central Balkan diffusion main route
(highlighted in \hyperref[fig:network]{Figure 17.5}) we analysed the
temporal evolution of their diffusion coefficients
(\hyperref[fig:trajectory]{Figure 17.6}). A similar pattern is visible
in all three regions and all diffusion coefficients: \(D\) starts at
zero, then rapidly rises to a marked peak and slowly decays
asymptotically to an intermediate value. This behavior is a consequence
of the local influence and its difference to adjacent regions. Initially
the influence difference is zero, because all regions have similar
technology and population. As soon as one region innovates (or receives
via diffusion technology and population from one of its neighbors),
population and technology increase, and so does the influence difference
to all other neighbors. With an increase in influence and diffusion
coefficient, \index{demic diffusion}demic and
\index{cultural diffusion}cultural diffusion to neighbors decrease the
influence differences. Relative proportions among the diffusion
coefficients of one region to all its neighbors are constant and
attributed to the geographical setting.

The time evolution of the diffusion coefficient plotted in
\hyperref[fig:trajectory]{Figure 17.6} reflects the population
statistics for advancing Neolithic technology: Early farming appears to
be associated to a rapid increase in population, this on a
supra-regional scale~{[}39{]}~{[}40{]}. At the regional level, the
diffusion coefficient lags the onset of farming by several hundred
years. This lag is also empirically reflected in the data set of the
western LBK~{[}41{]}. Any pioneering farming society seems to have
followed more or less the same general population trajectory with a
gradual increase over several centuries, followed by a sudden
rise-and-decline. The causes for this general pattern are yet unclear,
but may have to be sought more in social behavior patterns rather than
purely economic or environmental determinants~{[}42{]}.

\section{Conclusion}\label{conclusion}

It has been long evident, that the Neolithic ``Revolution'' is not a
single event, but heterogeneous in space and time. Statistical models
for understanding the diffusion processes, however, have so far assumed
that a physical model of Fickian diffusion can be applied to the pattern
of the emergence of farming and pastoralism using constant diffusion
coefficients. Relaxing this constraint, and reformulating the
diffusivity as a function of influence differences between regions,
demonstrates how diffusivity varies in space and time.

When results using this variable correlation coefficient (\(D\)) are
compared to empirical archaeological data, they represent the dynamics
on a continental scale and on the regional scale for many regions well,
but not for all: The impetus of the Neolithic in Greece and the Balkans
is well represented, also in southeastern Central Europe. The emergence
and the expansion of the Central European LBK shows, however, a too
early expansion in the model, whereas the stagnation following the
initial expansion is again very well represented.

Divergence between the mathematical model and the empirical findings
provided by archaeology is unsurprising and expected, because human
societies behave in much more complex ways than are described in the
highly aggregated and simplified model. Individuals may have chosen to
act independent of the social and environmental context and against
rational maximization of benefits. Rather than perfectly capturing each
regional diffusion event, the mathematical model serves as a null
hypothesis which is broadly consistent with the archaeologically
reconstructed picture, and against which individual decisions can be
assessed. In this respect, the simple model helps to disentangle in
complex histories general forcing agents and individual choices.

\textbf{References}

{[}1{]}V. G. Childe, \emph{Dawn Of European Civilization}, 1st ed.
(Routledge {[}reprinted 2005{]}, 1925), p. 256.

{[}2{]}N. Goring-Morris and A. Belfer-Cohen, Current Anthropology
\textbf{52}, (2011).

{[}3{]}G. Barker, \emph{The Agricultural Revolution in Prehistory: Why
Did Foragers Become Farmers?} (Oxford University Press, Oxford, United
Kingdom, 2006), p. 616.

{[}4{]}W. F. Ruddiman, D. Q. Fuller, J. E. Kutzbach, P. C. Tzedakis, J.
O. Kaplan, E. C. Ellis, S. J. Vavrus, C. N. Roberts, R. Fyfe, F. He, C.
Lemmen, and J. Woodbridge, Reviews of Geophysics \textbf{54}, 93 (2016).

{[}5{]}P. Bellwood, \emph{First Farmers: The Origins of Agricultural
Societies}, 1st ed. (Wiley-Blackwell, Malden, 2004), p. 384.

{[}6{]}E. M. Gallagher, S. J. Shennan, and M. G. Thomas, Proceedings of
the National Academy of Sciences \textbf{112}, 201511870 (2015).

{[}7{]}J. G. D. Clark, Proceedings of the Prehistoric Society
\textbf{31}, 57 (1965).

{[}8{]}M.-j. Gaillard, S. Sugita, F. Mazier, J. O. Kaplan, A.-K.
Trondman, A. Brostr{ö}m, T. Hickler, E. Kjellstr{ö}m, P. Kune{š}, C.
Lemmen, J. Olofsson, B. Smith, and G. Strandberg, Climate of the Past
Discussions \textbf{6}, 307 (2010).

{[}9{]}H. J. Zahid, E. Robinson, and R. L. Kelly, Proceedings of the
National Academy of Sciences 201517650 (2015).

{[}10{]}B. G. Purzycki, C. Apicella, Q. D. Atkinson, E. Cohen, R. A.
McNamara, A. K. Willard, D. Xygalatas, A. Norenzayan, and J. Henrich,
Nature \textbf{530}, 327 (2016).

{[}11{]}J.-P. Bocquet-Appel, S. Naji, M. {Vander Linden}, and J.
Kozlowski, Journal of Archaeological Science \textbf{39}, 531 (2012).

{[}12{]}F. Silva, J. Steele, K. Gibbs, and P. Jordan, Radiocarbon
\textbf{56}, 723 (2014).

{[}13{]}K. Manning, S. S. Downey, S. Colledge, J. Conolly, B. Stopp, K.
Dobney, and S. Shennan, Antiquity \textbf{87}, 1046 (2013).

{[}14{]}A. Bogaard, \emph{Neolithic Farming in Central Europe : An
Archaeobotanical Study of Crop Husbandry Practices} (Routledge, 2004),
p. 210.

{[}15{]}W. Schier, Praehistorische Zeitschrift \textbf{84}, 15 (2009).

{[}16{]}F.-X. Ricaut, Advances in Anthropology \textbf{02}, 14 (2012).

{[}17{]}W. Haak, I. Lazaridis, N. Patterson, N. Rohland, S. Mallick, B.
Llamas, G. Brandt, S. Nordenfelt, E. Harney, K. Stewardson, Q. Fu, A.
Mittnik, E. B{á}nffy, C. Economou, M. Francken, S. Friederich, R. G.
Pena, F. Hallgren, V. Khartanovich, A. Khokhlov, M. Kunst, P. Kuznetsov,
H. Meller, O. Mochalov, V. Moiseyev, N. Nicklisch, S. L. Pichler, R.
Risch, M. a. {Rojo Guerra}, C. Roth, A. Sz{é}cs{é}nyi-Nagy, J. Wahl, M.
Meyer, J. Krause, D. Brown, D. Anthony, A. Cooper, K. W. Alt, and D.
Reich, Nature \textbf{522}, 207 (2015).

{[}18{]}G. Brandt, A. Sz{é}cs{é}nyi-Nagy, C. Roth, K. W. Alt, and W.
Haak, Journal of Human Evolution \textbf{79}, 73 (2015).

{[}19{]}Z. Hofmanov{á}, S. Kreutzer, G. Hellenthal, C. Sell, Y.
Diekmann, D. {D{í}ez del Molino}, L. van Dorp, S. L{ó}pez, A.
Kousathanas, V. Link, K. Kirsanow, L. M. Cassidy, R. Martiniano, M.
Strobel, A. Scheu, K. Kotsakis, P. Halstead, S. Triantaphyllou, N.
Kyparissi-Apostolika, D.-C. Urem-Kotsou, C. Ziota, F. Adaktylou, S.
Gopalan, D. M. Bobo, L. Winkelbach, J. Bl{ö}cher, M. Unterl{ä}nder, C.
Leuenberger, Ç. {Ç}ilingiroğlu, B. Horejs, F. Gerritsen, S. Shennan, D.
G. Bradley, M. Currat, K. Veeramah, D. Wegmann, M. G. Thomas, C.
Papageorgopoulou, and J. Burger, Proceedings of the National Academy of
Sciences of the United States of America (2016).

{[}20{]}A. Ammerman and L. {Cavalli Sforza}, \emph{The Neolithic
Transition and the Population Genetics of Europe.} (Princeton
University, Princeton, NJ, 1984).

{[}21{]}J. Fort, Proceedings of the National Academy of Sciences of the
United States of America \textbf{109}, 18669 (2012).

{[}22{]}C. Lemmen, D. Gronenborn, and K. W. Wirtz, Journal of
Archaeological Science \textbf{38}, 3459 (2011).

{[}23{]}C. Lemmen, Archaeology, Ethnology and Anthropology of Eurasia
\textbf{41}, 48 (2013).

{[}24{]}C. Lemmen and A. Khan, in \emph{Climates, Landscapes, and
Civilizations}, edited by L. Giosan, D. Q. Fuller, K. Nicoll, R. K.
Flad, and P. D. Clift (American Geophysical Union, Washington, 2012),
pp. 107--114.

{[}25{]}K. W. Wirtz and C. Lemmen, Climatic Change \textbf{59}, 333
(2003).

{[}26{]}B. Arıkan, Journal of Archaeological Science \textbf{43}, 38
(2014).

{[}27{]}E. Boserup, \emph{Population and Technological Change}
(University of Chicago Press, 1981), p. 255.

{[}28{]}J. A. J. Metz, R. M. Nisbet, and S. A. H. Geritz, Trends in
Ecology and Evolution \textbf{7}, 198 (1992).

{[}29{]}R. A. Fisher, \emph{The Genetical Theory of Natural Selection}
(Dover, Oxford, 1930).

{[}30{]}K. W. Wirtz and B. Eckhardt, Ecological Modelling \textbf{92},
33 (1996).

{[}31{]}K. W. Wirtz, \emph{Modellierung von Anpassungsvorgängen in der
belebten Natur} (Universität Kassel, 1996), p. 216.

{[}32{]}C. Lemmen, Documenta Praehistorica \textbf{XLII}, 93 (2015).

{[}33{]}K. Davison, P. M. Dolukhanov, G. R. Sarson, and A. Shukurov,
Journal of Archaeological Science \textbf{33}, 641 (2006).

{[}34{]}R. Pinhasi, J. Fort, and A. J. Ammerman, Public Library of
Science Biology \textbf{3}, (2005).

{[}35{]}A. W. R. Whittle, F. M. A. Healy, and A. Bayliss,
\emph{Gathering Time: Dating the Early Neolithic Enclosures of Southern
Britain and Ireland} (Oxbow Books, 2011).

{[}36{]}C. Lemmen and K. W. Wirtz, Journal of Archaeological Science
\textbf{51}, 65 (2014).

{[}37{]}C. Jeunesse and S. {Van Willigen}, in \emph{The Spread of the
Neolithic to Central Europe} (Mainz, Germany, 2010), pp. 569--605.

{[}38{]}M. Rivollat, H. R{é}veillas, F. Mendisco, M.-H. Pemonge, P.
Justeau, C. Couture, P. Lefranc, C. F{é}liu, and M.-F. Deguilloux,
American Journal of Physical Anthropology \textbf{161}, 522 (2016).

{[}39{]}J.-P. Bocquet-Appel, Science \textbf{333}, 560 (2011).

{[}40{]}S. J. Shennan, S. S. Downey, A. Timpson, K. Edinborough, S.
Colledge, T. Kerig, K. Manning, and M. G. Thomas, Nature Communications
\textbf{4}, 1 (2013).

{[}41{]}D. Gronenborn, H.-C. Strien, S. Dietrich, and F. Sirocko,
Journal of Archaeological Science \textbf{51}, 73 (2014).

{[}42{]}D. Gronenborn, H.-C. Strien, and C. Lemmen, Quaternary
International, in press (2017).

\printindex

\end{document}